\documentclass[a4paper,twocolumn,superscriptaddress,11pt,accepted=2019-07-02]{quantumarticle}

\usepackage{graphicx,epsfig}
\usepackage{color}
\usepackage[usenames,dvipsnames]{xcolor}

\usepackage{amsmath,amssymb, amsthm, amsfonts}
\usepackage{dsfont} 
\usepackage{xspace}
\usepackage{xcolor}
\usepackage{changes}
\usepackage{verbatim}
\usepackage{mathtools}

\usepackage[ocgcolorlinks,colorlinks=true,linkcolor=blue,citecolor=red]{hyperref}

\usepackage[sort&compress,numbers]{natbib}

\newcommand{\one}{\mathds{1}}

\newcommand{\bb}[0]{\begin{eqnarray}}
\newcommand{\ee}[0]{\end{eqnarray}}
\newcommand{\ket}[1]{\vert #1\rangle}
\newcommand{\bra}[1]{\langle#1\vert}
\newcommand{\moy}[1]{\ensuremath{\langle #1\rangle}\xspace}
\newcommand{\uu}{{\uparrow \uparrow}}
\newcommand{\dd}{{\downarrow \downarrow}}
\newcommand{\ud}{{\uparrow \downarrow}}
\newcommand{\du}{{\downarrow \uparrow}}
\newcommand{\kuu}{{\ket{\!\!\uparrow \uparrow}}}
\newcommand{\kdd}{{\ket{\!\!\downarrow \downarrow}}}
\newcommand{\kud}{{\ket{\!\!\uparrow \downarrow}}}
\newcommand{\kdu}{{\ket{\!\!\downarrow \uparrow}}}
\newcommand{\buu}{{\bra{\uparrow \uparrow\!\!}}}
\newcommand{\bdd}{{\bra{\downarrow \downarrow\!\!}}}

\newcommand{\s}{{\cal S}}

\begin{document}

\title{Single-shot energetic-based estimator for entanglement in a half-parity measurement setup}

\author{Cyril Elouard}
\affiliation{Department of Physics and Astronomy, University of Rochester, Rochester, NY 14627, USA}
\affiliation{CNRS and Universit\'e Grenoble Alpes, Institut N\'eel, F-38042 Grenoble, France}

\author{Alexia Auff\`eves}
\affiliation{CNRS and Universit\'e Grenoble Alpes, Institut N\'eel, F-38042 Grenoble, France}

\author{G\'eraldine Haack}
\email{geraldine.haack@unige.ch}
\affiliation{Universit\'e de Gen\`eve, Department of Applied Physics, Chemin de Pinchat 22, CH-1211 Gen\`eve 4, Switzerland}

\date{\today}

\begin{abstract}
Producing and certifying entanglement between distant qubits is a highly desirable skill for quantum information technologies. Here we propose a new strategy to monitor and characterize entanglement genesis in a half parity measurement setup, that relies on the continuous readout of an energetic observable which is the half-parity observable itself. Based on a quantum-trajectory approach, we theoretically analyze the statistics of energetic fluctuations for a pair of continuously monitored qubits. We quantitatively relate these energetic fluctuations to the rate of entanglement produced between the qubits, and build an energetic-based estimator to assess the presence of entanglement in the circuit. Remarkably, this estimator is valid at the single-trajectory level and shows to be robust against finite detection efficiency. Our work paves the road towards a fundamental understanding of the stochastic energetic processes associated with entanglement genesis, and opens new perspectives for witnessing quantum correlations thanks to quantum thermodynamic quantities.
\end{abstract}

\maketitle

\section{Introduction} 
\label{Sec:intro}

Entanglement is a cornerstone of quantum information technologies: Entangled pairs of qubits have for long been a resource for secure quantum key distribution \cite{Ekert91}, quantum teleportation \cite{Bennett93}, quantum repeaters for long distance quantum communication \cite{Briegel98}, while large scale entanglement can be exploited through cluster states to perform measurement-based quantum computing \cite{Raussendorf01}. %The creation of quantum correlations in the presence of a thermal bath has some intrinsic work cost that has been analyzed in the context of resource theories \cite{Huber15,Bruschi15, Friis16}.

From a practical point of view, entangled pairs of qubits can be produced, e.g. by using quantum gates based on non-linear interactions \cite{Nielsen}, or performing measurement-based protocols of the parity observable. The latter is known to induce entanglement between two qubits initially in a separable state and has been extensively investigated in the last decade within various physical systems. Proposals have been made considering superconducting qubits jointly measured by a cavity mode \cite{Lalumiere10, Tornberg10, Chantasri16, Royer18} and semiconductor quantum dots jointly measured by a quantum point contact \cite{Trauzettel06, Williams08} or by an electronic Mach-Zehnder interferometer in quantum transport experiments \cite{Haack10, Meyer14}. All these works have derived the specific conditions under which these setups can be operated as parity meters. Those conditions include having a fine tuning of the coupling parameters between the qubits and the detector such that only the parity degree of freedom of the qubits is measured. In addition, having each of the two qubits initially in a maximal coherent superposition state is required to generate maximally entangled states. In particular, Ref. \cite{Williams08} investigated the stochastic generation of entanglement from a weak continuous measurement of the parity operator, putting forward the measurement-induced entanglement genesis. Since then, measurement-induced entanglement has eventually been implemented within circuit QED experiments \cite{Riste13, Roch14, Chantasri16}. Interestingly, these platforms also provide the technological know-how to access the quantum trajectories of individual quantum systems both subject to local measurements \cite{Weber14, Ficheux17, Naghiloo18} or to joint-measurements \cite{Roch14, Chantasri16}. Hence, Refs. \cite{Roch14, Chantasri16} not only implemented a parity-measurement based protocol onto two qubits, but could also access the stochastic trajectories followed by the joint state of the qubits along the entanglement generation process.

Recently, it was shown that the measurements allowing to reconstruct the pure state trajectories of the monitored systems are associated with energetic fluctuations of genuinely quantum origin called quantum heat. These energetic fluctuations can be turned into work in various protocols \cite{Elouard17b, Elouard18, Buffoni18, Ding18} and have been related to entropy production of quantum origin \cite{Elouard17, Manzano15} and provide new merit criteria to assess the performances of a feedback loop \cite{Alonso16, Naghiloo18}.  So far, a thermodynamic analysis of entanglement genesis based on stochastic trajectories has remained elusive.

In this work, we theoretically analyze the statistics of energetic fluctuations for a pair of continuously monitored half spins subject to a half-parity measurement. This is achieved within the framework of stochastic quantum thermodynamics presented in \cite{Elouard17,Elouardchapter} and it allows us to derive and highlight for the first time energetic signatures associated with measurement-induced entanglement genesis. We quantitatively relate these energetic quantum fluctuations to the rate of entanglement produced between the qubits. We then exploit our results to propose a new practical application of these energetic fluctuations by building an energetic-based estimator to attest the presence of entanglement in the circuit. Remarkably, this latter quantity holds at the single trajectory level and does not rely on the measurement record itself.

%Here we propose a new practical application of these energetic fluctuations, based on the analysis of the half parity measurement protocol in the framework of quantum stochastic thermodynamics. Based on a quantum-trajectory approach, we theoretically analyze the statistics of energetic fluctuations for a pair of continuously monitored half spins. We quantitatively relate these quantum fluctuations to the rate of entanglement produced between the qubits, and build an energetic-based estimator to attest the presence of entanglement in the circuit. Remarkably, this analysis holds at the single trajectory level.
The paper is organized as follows. In Sec.~\ref{sec2}, we present our system and recall the basics of a half-parity measurement protocol. 
In Sec.~\ref{sec3}, we relate parity measurement and energy measurement, and we introduce the quantum energetic fluctuations associated with the quantum heat. We define and compute the stochastic energetic quantities involved in the continuous measurement case. In Sec.~\ref{sec4}, we relate the fluctuations of quantum heat to the rate of entanglement induced between the qubits. This one is investigated through the time-derivative of the concurrence, a monotone measure for two-qubit entanglement. We derive upper and lower bounds for the rate of entanglement genesis, which we exploit in Sec.~\ref{sec5} to build an estimator assessing the presence of entanglement. The latter does not depend explicitly on the measurement record, it is solely based on energetic quantities and is valid at the single-trajectory level. We also investigate its robustness in presence of finite detection efficiency.

\begin{figure}[t]
\begin{center}
\includegraphics[width=\linewidth]{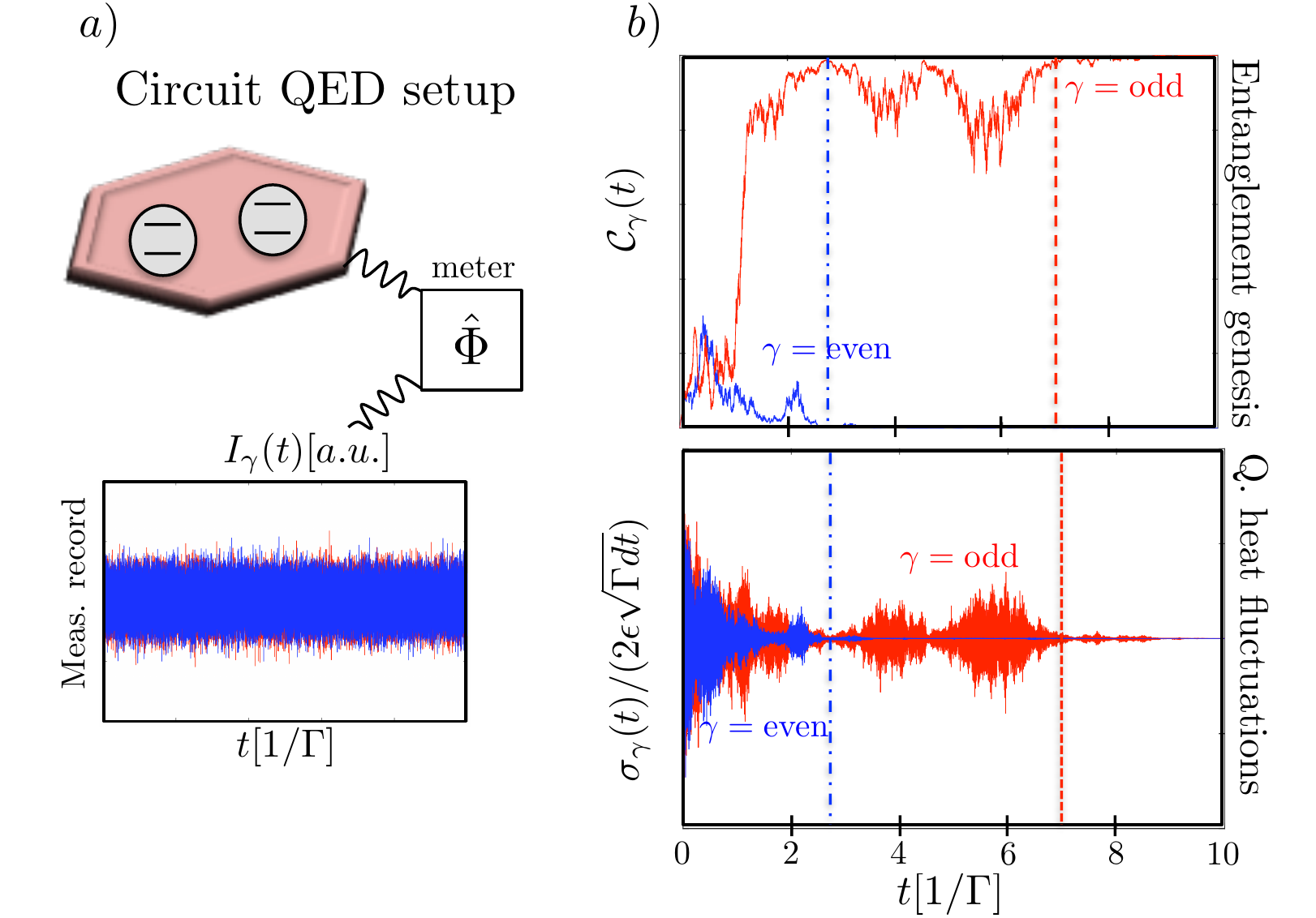}
\caption{Model and motivation. a) Weak continuous half-parity measurement. Two qubits are weakly coupled to a meter able to measure continuously the joint observable $\hat{\Phi}$, for instance in circuit QED setups, see Refs. \cite{Riste13, Roch14}. The corresponding measurement record $I_\gamma(t)$ is stochastic. Two realizations are shown: trajectory leading the qubits into a maximally entangled state (red) or into a product state (blue). b) After post-selection, time-dependent single-trajectory concurrence and single-trajectory quantum heat fluctuations. Vertical dashed lines highlight similar times for the concurrence and the quantum heat fluctuations to reach their long-time limit value. In this work we derive formal relations between those quantities and exploit them to derive an energetic-based estimator, an alternative to quantum state tomography.}
\label{fig:setup}
\end{center}
\end{figure}

%\section{Results}
\section{Model} 
\label{sec2}

\textit{System--} The system is made of two two-level systems (qubits) with identical energy splitting $\epsilon$ described by their Hamiltonian $\hat H_{\s}$ 
\bb
\label{eq:Hqb}
\hat H_{\s} =\epsilon \left( \hat \sigma_z^{(1)}\otimes \one + \one \otimes \hat \sigma_z^{(2)}\right) \,,%= \epsilon \, \hat \Phi\,.
\ee
where $\sigma_z^{(i)}$ denotes the z-Pauli matrix for qubit $i$. We assume the two qubits to be initially in a separable state, and more specifically in a maximal superposition state:
\bb
\label{eq:psi_0}
\ket{\psi(0)} = \frac{1}{2} \left(\ket{\!\uparrow} + \ket{\!\downarrow}  \right) \otimes \left(\ket{\!\uparrow} + \ket{\!\downarrow}  \right)\,.
\ee
This choice is motivated by previous works showing that this state belongs to the set of optimal states that lead to maximally entangled final states (i.e. Bell states) when the qubits are subject to a (half-) parity measurement \cite{Haack10}. Hence, this choice of initial state will allow us to investigate the energetic signatures associated to the generation of entanglement in optimal conditions.
In addition, the qubits are subject to a weak continuous measurement of the joint observable $\hat{\Phi}$, that implements a half-parity measurement, see Fig. \ref{fig:setup}. Within the two-qubit computational basis $\{ \kuu, \kud, \kdu, \kdd \}$, the collective operator $\hat{\Phi} = \sum_{i=1,2} \sigma_z^{(i)}$ is defined by $\kuu\buu- \kdd\bdd$ and has three eigenvalues $\pm1$ and $0$ with eigenstates $\kuu$, $\kdd$ and $(\kud \pm \kdu)/\sqrt{2}$ respectively. The eigenvalue 0 is degenerate, implying that this outcome does not allow one to distinguish between the two odd states $\{ \kud, \kdu\}$. Consequently, when outcome $0$ is obtained from the measurement, the two qubits are driven into a coherent superposition of those states, which leads to entanglement. On the contrary, the measurement of the eigenvalues $\pm 1$ allows to distinguish the two even states, leaving the qubits in a product state. Hence, the half-parity measurement presents the specificity of producing in a probabilistic way entangled and product states. It is clear from the expression of $\hat{\Phi} $ that its measurement does not put the qubits into an entangled state (spanned by the odd states) if these ones are not in a coherent superposition initially. Although the forms of $\hat{\Phi}$ and of the parity operator $\hat{P} = \sigma_z^{(1)} \otimes \sigma_z^{(2)}$ seem to be specific, these joint operators are the only ones up to local unitary operations that can generate correlations between two qubits. This explains the high interest in the past years for this parity measurement protocol within solid-state setups, also known within the quantum information community as a Bell measurement. This one was for instance proposed and realized for quantum teleportation \cite{Bennett93, Riebe04}. \\%Remarkably, the half-parity observable also serves as energy filter within our model: $\hat{\Phi} \propto \hat{H}_S$, both being related by the energy gap of the qubits $\epsilon$ (see Eq. \eqref{eq:Hqb}). Hence, the half-parity measurement, in addition to generating entanglement, also provides a direct access to the energetics of the qubits along the quantum trajectories generated by the weak measurement of the corresponding $\hat{\Phi}$. This allows us to distinguish quantum trajectories associated with entanglement genesis or not using post-selection, and to compare their energetic properties.\\

\textit{Quantum trajectories--} Assuming a weak continuous measurement of $\hat{\Phi}$, each realization of the measurement procedure is associated to a quantum stochastic trajectory followed by the qubits and labelled $\gamma$ in the rest of the manuscript \cite{Wiseman96, Jacobs06}. We assume the initial state of the qubits is a known pure state, see Eq.\eqref{eq:psi_0}, such that the trajectory $\gamma$ is made of a sequence of pure states $\{\ket{\psi_\gamma(t)}\}$. To each trajectory $\gamma$ corresponds a stochastic measurement record $I_\gamma(t)$:
\bb
\label{eq:meas_I}
I_\gamma(t) = \moy{\hat\Phi(t)}_\gamma + \frac{dW_\gamma(t)}{2\sqrt{\Gamma}dt}\,.
\ee
Here, $ \moy{\hat\Phi(t)}_\gamma= \bra{\psi_\gamma(t)}\hat\Phi\ket{\psi_\gamma(t)}$ is the expectation value of the half-parity operator w.r.t. the state $\ket{\psi_\gamma(t)}$, and $\Gamma$ corresponds to the detector measurement rate, i.e. the rate at which one is able to distinguish the measurement outcomes from the detector's shot noise \cite{Pilgram02, Korotkov99, Haack10, Meyer14}. The infinitesimal Wiener increment $dW_\gamma(t)$ is a stochastic variable characterized by a zero average and variance $dt$, i.e. $\langle\!\langle dW_\gamma \rangle\!\rangle_t =0$ and $\langle\!\langle dW_\gamma^2 \rangle\!\rangle_t= dt$, where $\langle\!\langle \cdot \rangle\!\rangle_t$ denotes the average over all realizations of the measurement during the time interval $[0, t]$.  Note that throughout this work, we use Ito's convention for stochastic differential calculus \cite{Gardiner85, Gillespie96}. This infinitesimal Wiener increment encodes Gaussian fluctuations of the measurement record $I_\gamma(t)$ around its expectation value $\moy{\hat\Phi(t)}_\gamma$ and therefore captures the detector's shot noise in the weak coupling limit \cite{Korotkov99, Jacobs06}. 
Based upon the knowledge of $I_\gamma$, the conditional dynamics of the two qubits subject to the weak measurement of the half-parity observable $\hat{\Phi}$ is captured by the stochastic Schr\"odinger equation
\bb
\label{SSE}
d\ket{\psi_\gamma(t)} &=& \left[-i \hat H_{\s} dt- \frac{\Gamma}{2} dt (\hat{\Phi} - \moy{\hat\Phi(t)}_\gamma)^2\right. \nonumber\\
&& \left. \;+ \sqrt{\Gamma} \, dW_\gamma(t)(\hat \Phi -  \moy{\hat\Phi(t)}_\gamma)\right] \ket{\psi_\gamma(t)} .
\ee 

\begin{figure}[t!]
\begin{center}
\includegraphics[width=0.9\linewidth]{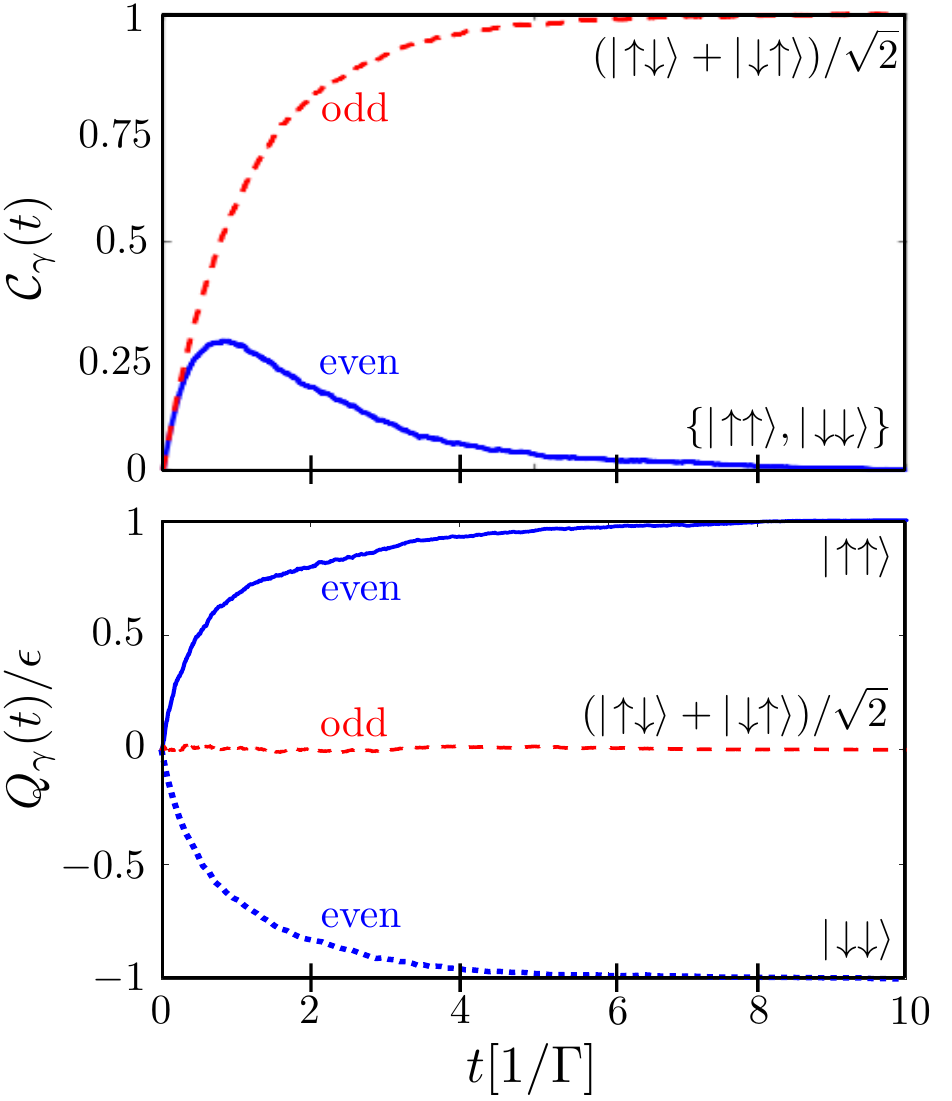}
\end{center}
\caption{Post-selected average concurrence $\mathcal{C}_\gamma$ and quantum heat $Q_\gamma$ (as a function of time i.u. of $[1/\Gamma]$). Red dashed curves are obtained by averaging over all \textit{odd} trajectories, \textit{i.e.} that lead the qubits into the maximally entangled state $(\ket{\ud} + \ket{\du})/\sqrt{2}$. Blue (full, dashed) curves are obtained by averaging over all \textit{even} trajectories, \textit{i.e.} that lead the qubits into a product state $(\ket{\uu}, \ket{\dd})$. The half-parity measurement induces entanglement, but also serves as energy filter. When averaged over all trajectories (even and odd), $\langle\!\langle Q_\gamma \rangle\!\rangle =0$ as expected for a QND-measurement $[\hat{H}_S, \hat{\Phi}]=0$. Total number of simulated quantum trajectories: 800}
\label{fig:av_thermo}
\end{figure}

Using $\hat H_{\s} = \epsilon \hat\Phi$, see Eq. \eqref{eq:Hqb}, one can solve analytically Eq.\eqref{SSE} at any time $t$ as a function of the stochastic measurement record:
\bb
\label{eq:state}
\ket{\psi(J_\gamma,t)} = \frac{1}{N_\gamma(t)}\left(e^{(-i\epsilon  + 2\Gamma J_\gamma \, )\hat \Phi t - \Gamma  \hat\Phi^2 t} \right)\ket{\psi(0)}\,,
\ee
with $N_\gamma(t)$ the time-dependent normalization factor $N_\gamma(t)=  \sqrt{(1+ e^{-2 \Gamma t} \cosh(4 \Gamma J_\gamma t ))/2}$. Following Ref. \cite{Jacobs06}, this solution is expressed in terms of the measurement outcome $J_\gamma = 1/t \, \int_0^t I_\gamma(\tau) \, d\tau $:
\bb\label{Jgamma}
J_\gamma(t) =  \frac{1}{t} \int_0^t\moy{\hat\Phi(\tau)}_\gamma \, d\tau\, + \frac{W_\gamma(t)}{2\sqrt{\Gamma} t}\,,
\ee 
with $W_\gamma(t) = \int_0^t  dW_\gamma(t')$ a Gaussian random variable with mean zero and variance $t$. The time integral corresponds to a finite resolution time, typically set by experimental constraints. The probability distribution of $J_\gamma(t)$ is a sum of three Gaussian functions of variance $1/4\Gamma t$, each peaked around one of the eigenvalues $\{0, \pm 1\}$ of the measured observable $\hat \Phi$:
\bb
\label{eq:proba_dis}
P(J_\gamma,t) = \frac{1}{3} \sum_{j=-1}^1 \sqrt{\frac{2 \Gamma t}{\pi}} e^{- (J_\gamma-j)^2/ (2 \tilde{\sigma}_0^2)}\,,
\ee
with the variance $\tilde{\sigma}_0 \equiv \sqrt{\Gamma t}$ setting the measurement strength. In the long time limit $t\gg (4\Gamma)^{-1}$, $J_\gamma(t)$ only takes one of three possible outcomes which are the eigenvalues of the half-parity measurement operator: $J_0 \equiv 0$, $J_{\pm1}\equiv \pm 1$. At those long times, the measurement becomes projective. In the following of the work, we choose to label the different trajectories with their long-time value of $J_\gamma$; this defines three subsets of trajectories, leading the qubits either in an entangled state ($J_\gamma = 0$) or in a product state ($J_\gamma = \pm1$).\\

\textit{Measure of entanglement--} We quantify the presence of entanglement at any time $t$ using the concurrence, a monotone measure for two-qubit entangled states \cite{Wootters98}. The qubits being in a pure state at each instant of time along their trajectory $\gamma$, we can make use of a simpler definition $\mathcal{C}_\gamma(t) = \text{max}\left\{ 0, 2\vert ad - bc \vert \right\}$ where $a,b,c,d$ are the amplitudes of $\ket{\psi_\gamma(t)}$ with respect to the computational states $\kuu, \kud,\kdu, \kdd$ respectively. Inserting Eq.\eqref{eq:state}, we get a stochastic concurrence which depends on the measurement outcome $J_\gamma$:
\bb
\label{eq:C}
\mathcal{C}_\gamma(t) &=&  \frac{1-e^{- 2 \Gamma t}}{1+ e^{-2 \Gamma t} \cosh(4 \Gamma J_\gamma t)}.
\ee

Figure~\ref{fig:av_thermo}\,a) shows the time evolution of the average post-selected concurrence $\mathcal{C}_\gamma$ according to the three subsets of trajectories, labeled by the final outcomes $J_\gamma \to \{J_0, J_{\pm1}\}$. The average is made over all realizations that belong to a given subset. At long times, $t > 6/\Gamma$, the concurrence can be directly obtained by replacing in Eq.\eqref{eq:C} $J_\gamma$ by the $\hat{\Phi}$-eigenvalues $0, \pm 1$ as the measurement becomes projective. Under optimal conditions as considered here (QND-measurement, no additional dephasing processes, ideal detection scheme), the concurrence reaches the value 1 when qubits are driven into the odd subspace, meaning that the qubits end up in a maximally entangled state. Non-ideal conditions encountered in experiments lead to a lower maximal value of concurrence, and eventually to a long-time state that is separable if dephasing processes are too important. This was for instance the case in Ref. \cite{Roch14}, and can be accounted for theoretically.\\

In this work, the goal is to establish fundamental links between energetic signatures and the generation of entanglement. This is why we focus below on the realization of weak continuous half-parity measurement under optimal conditions. Non-ideal conditions could also be accounted for, but would prevent us to draw clear conclusions on the origin of these energetic signatures.

\section{Half-parity measurement seen as an energy measurement (filter)}
\label{sec3}

It is remarkable that the half-parity observable also serves as energy filter within our model: $\hat{\Phi} \propto \hat{H}_S$, both being related by the energy gap of the qubits $\epsilon$ (see Eq. \eqref{eq:Hqb}). Hence, the half-parity measurement provides a direct access to the energetics of the qubits along the quantum trajectories generated by the weak measurement of $\hat{\Phi}$. The internal energy $U_\gamma(t)$ of the two qubits along a given trajectory $\gamma$ is given by:
\bb
\label{eq:U_def}
U_\gamma(t) = \bra{\psi_\gamma(t)} \hat H_{\s} \ket{\psi_\gamma(t)} = \epsilon  \bra{\psi_\gamma(t)} \hat{\Phi} \ket{\psi_\gamma(t)}\,.
\ee
Considering the initial state $\ket{\psi(0)}$ (see Eq.\eqref{eq:psi_0}), the initial internal energy $U(0)$ is zero and constitutes the energy reference. In the absence of driving, $\hat H_{\s}$ is time-independent. Hence no work is performed onto the qubits. In addition, there is no thermal reservoir involved in the problem so that a change in the internal energy of the qubits can only arise from the measurement process itself. This form of energy exchange has no classical counter-part \cite{Brandner15,Alonso16, Elouard17, Abdelkhalek16, PRL17, Yi17} and will be, in the line of \cite{Elouard17}, referred to as quantum heat and denoted $Q$ in this article. \\

\textit{Long times: projective energetic measurement--} At long times, $t > (6\Gamma)^{-1}$, the change of internal energy (associated here to a net quantum heat $Q$) takes three different values depending on the measurement outcomes:
\bb
\Delta U \equiv Q = \left\{\begin{array}{c}
0 \quad \text{for} \quad J_\gamma = 0 \\ 
\epsilon \quad \text{for} \quad J_\gamma = 1 \\ 
-\epsilon \quad \text{for} \quad J_\gamma = -1.
\end{array} \right.
\ee
When averaged over all trajectories, $\left\langle\!\left\langle \Delta U \right\rangle\!\right\rangle_\gamma = 0$, which equals the reference internal energy at initial time $U(0) = 0$. This equality follows from $[\hat{H}_S, \hat{\Phi}] = 0$ that characterizes a QND-measurement. Indeed, as both observables commute, there must be no change of internal energy on average and this must hold at all times. At long times, the weak continuous measurement of $\hat{\Phi}$ is equivalent to a projective \textit{energy} measurement, and this explains the long-time values of $\Delta U$ for the different subsets of trajectories, equal to the energy of the odd states $\{ \ket{\ud}, \ket{\du}\}$ and even states $\ket{\uu}$ and $\ket{\dd}$ respectively.\\

\textit{Intermediate times--} At arbitrary time $t$, the quantum heat exchange depends on the exact outcome $J_\gamma(t)$:
\bb
\label{eq:Q}
Q_{\gamma}(t) &=& U_\gamma(t) - U(0) \nonumber \\
&=& \epsilon \,  \frac{ e^{- 2 \Gamma t} \sinh(4\Gamma t J_\gamma)}{1+ e^{-2 \Gamma t} \cosh(4\Gamma t J_\gamma)\,.}
\ee
This quantum heat exchange $Q_{\gamma}(t)$, after post-selection, is plotted in Fig.~\ref{fig:av_thermo}~b) as a function of time. As for the concurrence, an average is made within each subset of trajectories defined by the long-time limit value of the measurement record $J_\gamma = 0, \pm 1$. Because the half-parity measurement amounts to an energetic measurement, an access to the internal energy is sufficient to determine whether the qubits are entangled or not. A record equal to 0 implies that the qubits ended up in the odd subspace, in a coherent superposition of the odd states $(\ket{\ud} + \ket{\du})\sqrt{2}$. 
%This result broadens the range of applications of the half-parity measurement to quantum thermodynamics: Introduced in previous works as a way to generate entangled state in a heralded way (the outcome is enough to know whether the final state is entangled), this first part demonstrates that it also gives access to the joint internal energy of the qubits.
This first part demonstrates that, on top of being a way to generate entangled state in a heralded way (the outcome is enough to know whether the final state is entangled) as demonstrated in previous works, the half-parity measurement also gives access to the joint internal energy of the qubits. Let us note that the rate at which the quantum heat exchange converges to its final value corresponds to the measurement-induced dephasing rates derived in previous works \cite{Haack10, Meyer14, Riste13,Roch14}.\\

\textit{Stochastic fluctuations at arbitrary times--} %In this work, we go beyond the investigation of the internal energy and the exchanged quantum heat by deriving their stochastic fluctuations and variance. As shown in the next section, this allows us to formally derive energetic bounds for the rate of entanglement formation, and to propose an estimator to assess the presence of entanglement based solely on energetic measurements.  
{To derive the fluctuations of the quantum heat,} let us first introduce the increment $\delta Q_\gamma(t)$ that corresponds to the stochastic infinitesimal variation of the internal energy $U_\gamma(t)$ between times $t$ and $t+dt$ along a trajectory $\gamma$. It is related to the total quantum heat exchange $Q_\gamma(t)$ up to time $t$ via:
\bb
\label{eq:Q_dQ}
Q_\gamma(t) = \int_0^t \delta Q_\gamma(t')\,,
\ee 
%Using Ito's rule of stochastic calculus, this infinitesimal heat exchange is defined as
and is defined as
\bb
\delta Q_\gamma(t) &\equiv& d\left(\langle \psi_\gamma(t) \vert \hat H_{\s}\vert \psi_\gamma(t) \rangle \right) \\
&=& \Big( d\langle \psi_\gamma(t) \vert \Big) \hat H_{\s}\vert \psi_\gamma(t) \rangle + \langle \psi_\gamma(t) \vert \hat H_{\s} \Big( d\vert \psi_\gamma(t) \rangle \Big) \nonumber \\
&& +  \Big( d\langle \psi_\gamma(t) \vert \Big) \hat H_{\s} \Big( d\vert \psi_\gamma(t) \rangle \Big) \label{eq:dQ}
\ee

Inserting Eqs.~\eqref{eq:Hqb} and \eqref{eq:state} into Eq.~\eqref{eq:dQ} and expanding the last term up to the first order in $dt$, we obtain
\bb
\label{eq:dQ1}
\delta Q_\gamma(t) &=& 2 \epsilon \sqrt{\Gamma} dW_\gamma(t) \left( 4 P_{\uparrow \uparrow} P_{\downarrow \downarrow} + P_o P_e \right) \nonumber \\
&\equiv& \delta Q^{(e)}_\gamma(t) + \delta Q^{(eo)}_\gamma(t)\,,
\ee
with
\bb
\delta Q^{(e)}_\gamma(t) &=& 8 \epsilon \sqrt{\Gamma} dW_\gamma(t) P_{\uparrow \uparrow} P_{\downarrow \downarrow} \label{eq:dQ2} \\
\delta Q^{(eo)}_\gamma(t) &=& 2 \epsilon \sqrt{\Gamma} dW_\gamma(t) P_e P_o \label{eq:dQ3}\,.
\ee
Here, we have introduced the populations $P_{ij}$ for the 4 two-qubit states defined as $P_{ij} \equiv P_{ij,\gamma}(t) = \vert \langle ij \vert \psi_\gamma(t)\rangle\vert^2 \quad i,j= \uparrow, \downarrow$
and we denote $P_e = P_{\uparrow \uparrow} + P_{\downarrow \downarrow}$  and $P_o = P_{\uparrow \downarrow} + P_{\downarrow \uparrow} $ the populations within the even and odd parity subspaces respectively. From those definitions, the products $P_{\uparrow \uparrow}P_{\downarrow \downarrow} $ and $P_e P_o$ correspond respectively to the squared coherences (off-diagonal elements in the two-qubit density matrix) within the even subspace and between the even and odd subspaces. Let us recall that a (half-) parity measurement generates entanglement by distinguishing the even states form the odd ones. It is therefore the loss of coherence between the two parity subspaces that brings the two qubits in a coherent superposition of odd states when outcome $J_\gamma = 0$ is obtained. Hence, we claim that the product $P_o P_e$ (equivalently $\delta Q^{(eo)}_\gamma(t)$) in Eq.\eqref{eq:dQ3} reflects entanglement genesis, and so does the total heat increment $\delta Q_\gamma(t)$. We demonstrate this claim in the following sections.

\section{Energetic bounds for the entanglement genesis rate}
\label{sec4}

To validate our claim, we define the standard deviation of the quantum heat increment between times $t$ and $t+ dt$ as
\bb
\sigma_\gamma(t) &=& \sqrt{\langle\!\langle \delta Q_\gamma^2(t) \rangle\!\rangle_{dt}} \nonumber\\
%\sigma_\gamma(t) &=& \sqrt{\left\langle\delta Q_\gamma(t)^2\right\rangle_{I_\gamma(t)}} \nonumber\\
&=& 2 \epsilon \sqrt{\Gamma dt}\, \frac{e^{-4 \Gamma t}+e^{-2 \Gamma t} \cosh( 4\Gamma t J_\gamma)}{\left( 1+ e^{-2 \Gamma t} \cosh(4\Gamma t J_\gamma) \right)^2} \,,\label{eq:s_gamma}
\ee
where we made use of $\langle\!\langle dW(t)^2 \rangle\!\rangle_{dt} = dt$ and $\langle\!\langle dW(t) \rangle\!\rangle_{dt} = 0$.
%The ensemble average $\moy{\cdot}_{I_\gamma(t)}$ is an average over the trajectories (or equivalently over the measurement record $I_\gamma(t)$), during the time interval $[t, t+dt]$, keeping the past records $\{I_\gamma(t')\}_{t'<t}$ fixed. This ensemble average allows us to get rid of the Wiener increment in Eqs.\eqref{eq:dQ2} and \eqref{eq:dQ3} as $\moy{dW^2_\gamma(t)}_{I_\gamma(t)}= dt$. 
For simplicity, we will work in the following with dimensionless quantities:
\bb
\label{eq:tilde}
\tilde{\sigma}_\gamma(t) = \frac{\sigma_\gamma(t)}{ 2 \epsilon \sqrt{\Gamma dt}} \quad \text{and} \quad \tilde{Q}_\gamma(t) &=& \frac{Q_\gamma(t)}{\epsilon}\,.
\ee
$\sqrt{\Gamma dt}$ refers to the variance of the Gaussian distribution of the measurement record $I_\gamma(t)$ during the discrete time interval $dt$, see the variance $\tilde{\sigma}_0$ defined in Eq. \eqref{eq:proba_dis}. Similarly to Eq. \eqref{eq:dQ1}, we define two contributions to the standard deviation of the stochastic quantum heat increment:
\bb
\tilde{\sigma}_\gamma(t) = \tilde{\sigma}_\gamma^{(eo)}(t)+ \tilde{\sigma}_\gamma^{(e)}(t)
\ee

\begin{figure}[h]
\begin{center}
\includegraphics[width=0.9\linewidth]{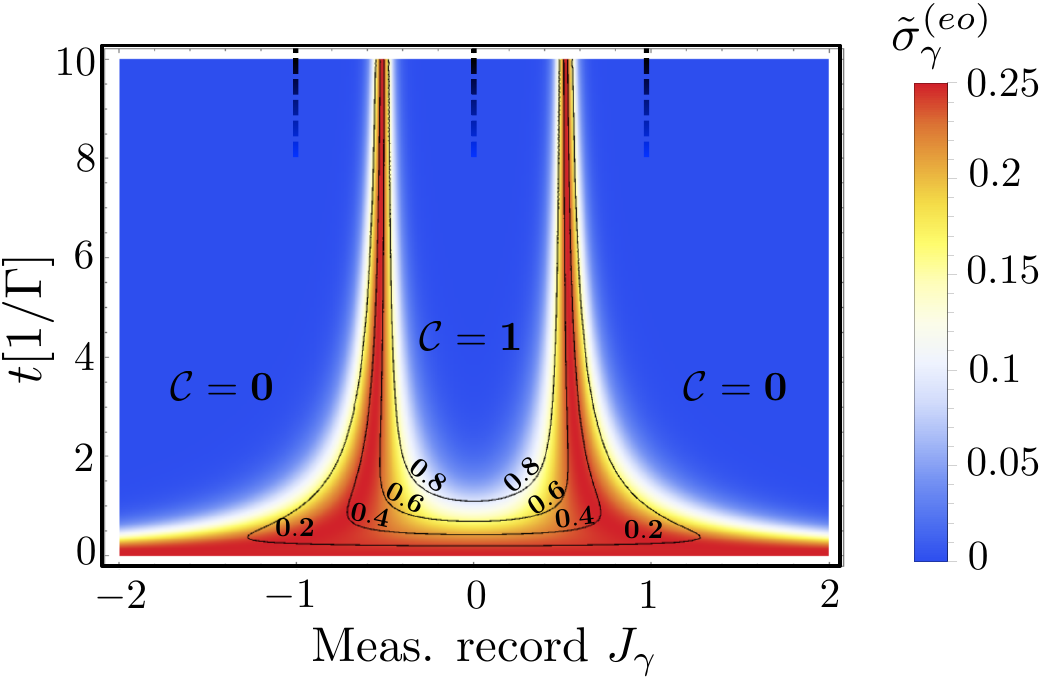}
\end{center}
\caption{Infinitesimal quantum heat fluctuations $\tilde{\sigma}_\gamma^{(eo)}$ as a function of the outcome $J_\gamma$ and the duration of the measurement $t$. The contour plot (solid black lines) corresponds to the lines of constant concurrence (from 0.2 to 0.8). Regions for $\mathcal{C} = 0,1$ correspond to the blue ones and are explicit on the figure. There exists a one to one correspondance between a finite derivative of the concurrence ($\mathcal{C}$ is changing) and finite fluctuations of the increment of quantum heat exchange. At long times, measurement outcome $J_\gamma$ converges to $0$ or $\pm1$, shown with dashed lines (not anymore continuous along the x-axis). Fluctuations are not present ($\tilde{\sigma}_{0, \pm1}^{(eo)}=0$)and $\mathcal{C} = \{1,0\}$ for $J_\gamma = \{0, \pm1\}$. %Dashed vertical lines indicate the three eigenvalues of $\hat\Phi$ corresponding to the three peaks in the distribution of $J_\gamma(t)$ at times $t>1/\Gamma$.
}\label{f:fig3}
\end{figure} 

Figure \ref{f:fig3} illustrates our claim. The contour plot corresponding to constant values of the concurrence is  superimposed onto the density plot of $\tilde{\sigma}^{(eo)}(t)$. It highlights that an increase in the concurrence is associated with non-zero fluctuations of $\delta Q^{(eo)}_\gamma(t)$. In contrast, the concurrence plateaus correspond to the areas where $\tilde{\sigma}^{(eo)}$ vanishes. This observation is supported by a strict equality between the concurrence variation (the concurrence derivative) and $\tilde{\sigma}_\gamma^{(eo)}$ for the trajectories leading the qubits into an entangled state $J_\gamma = 0$ (see App. \ref{appen:1} for detailed derivation):
\bb
\label{eq:ll}
\left.\frac{{d\mathcal{C}}_\gamma}{dt} \right\vert_{J_\gamma = 0} = 4 \Gamma \left.\tilde{\sigma}_\gamma^{(eo)} \right\vert_{J_\gamma = 0}\,.
\ee
%\deleted{This result demonstrates the importance of stochastic fluctuations within the context of stochastic quantum thermodynamics to access genuine quantum features at a fondamental level.} \deleted{Let us \deleted{emphasize} \gh{note} that the \deleted{long-time} equality Eq. \eqref{eq:ll}holds as the half-parity measurement then becomes projective.} 
Because it constitutes at the same time the source of entanglement and a direct access to the energy and its fluctuations, it follows that the energy fluctuations must be related to the creation of entanglement. However, the practical interest of Eq. \eqref{eq:ll} is limited as it would be impossible in a heat-sensitive measurement to distinguish the two contributions $\delta Q^{(eo)}_\gamma$ and $\delta Q^{(e)}_\gamma$.\\% and taking {$J_\gamma = 0$ has a true physical meaning} when the measurement becomes projective {at long times}. \\

Nevertheless, at intermediate times, we demonstrate that the concurrence derivative, that characterizes the rate at which entanglement is induced by the measurement, can be upper and lower bounded by energetic quantities solely. Starting from
\bb\label{eq:dCdt}
\frac{d \mathcal{C}_\gamma(t)}{dt} &=& 2 \Gamma \Bigg[ \frac{e^{-2 \Gamma t} \left( 1+ \cosh(4 \Gamma t J_\gamma)\right)}{(1+e^{-2 \Gamma t} \cosh(4 \Gamma t J_\gamma)^2}  \nonumber \\
&&  - \frac{2\mathcal{C}_\gamma (t) Q_\gamma(t)}{\epsilon} \left(\moy{\hat\Phi(t)}_\gamma + \frac{1}{2\sqrt{\Gamma}}\frac{d W_\gamma(t)}{dt} \right) \Bigg]\,,\nonumber\\
\ee
we then perform an an ensemble average over all trajectories (or equivalently over the measurement record $I_\gamma(t)$) during the time interval $[t, t+dt]$, keeping the past records $\{I_\gamma(t')\}_{t'<t}$ fixed. This interval $[t, t+ dt]$ is the one over which we investigate the fluctuations of the quantum heat increment $\delta Q_\gamma$, and this ensemble average $\langle \!\langle \cdot \rangle \! \rangle_{dt}$ corresponds to the one used in the definitions of the quantum heat increment fluctuations, see Eq.\eqref{eq:s_gamma}. We then have $\langle \!\langle dW_\gamma(t)/dt\rangle \! \rangle_{dt} = 0$ in Eq.\eqref{eq:dCdt} and

\bb
\label{eq:dC_eo}
\left\langle \!\!\!\left\langle \frac{d \mathcal{C}_\gamma(t)}{dt} \right\rangle \!\!\! \right\rangle_{dt} &=& 2 \Gamma \frac{e^{-2 \Gamma t} + e^{- 2 \Gamma t} \cosh(4 \Gamma t J_\gamma)}{(1+e^{-2 \Gamma t} \cosh(4 \Gamma t J_\gamma))^2} \nonumber \\
&&- 4 \Gamma \, \frac{Q_\gamma}{\epsilon} \, \mathcal{C}_\gamma  \moy{\hat\Phi(t)}_\gamma .
\ee
The first term on the r.h.s. can be upper and lower bounded with $\tilde{\sigma}_\gamma$ (see Append. \ref{appen:2}):
\bb
&&\left\langle \!\!\!\left\langle \frac{d \mathcal{C}_\gamma(t)}{dt} \right\rangle \!\!\! \right\rangle_{dt} \geq 2 \Gamma \left[ \tilde{\sigma}_\gamma(t) - 2 \tilde{Q}^2_\gamma(t) \mathcal{C}_\gamma(t) \right]  \label{eq:ineq1}\\
&&\left\langle \!\!\!\left\langle \frac{d \mathcal{C}_\gamma(t)}{dt} \right\rangle \!\!\! \right\rangle_{dt} \leq  2 \Gamma \left[2 \tilde{\sigma}_\gamma(t) - 2 \tilde{Q}^2_\gamma(t) \mathcal{C}_\gamma(t)\right] \label{eq:ineq2}
\ee

\begin{figure*}[t]
\begin{center}
\includegraphics[width=0.9\linewidth]{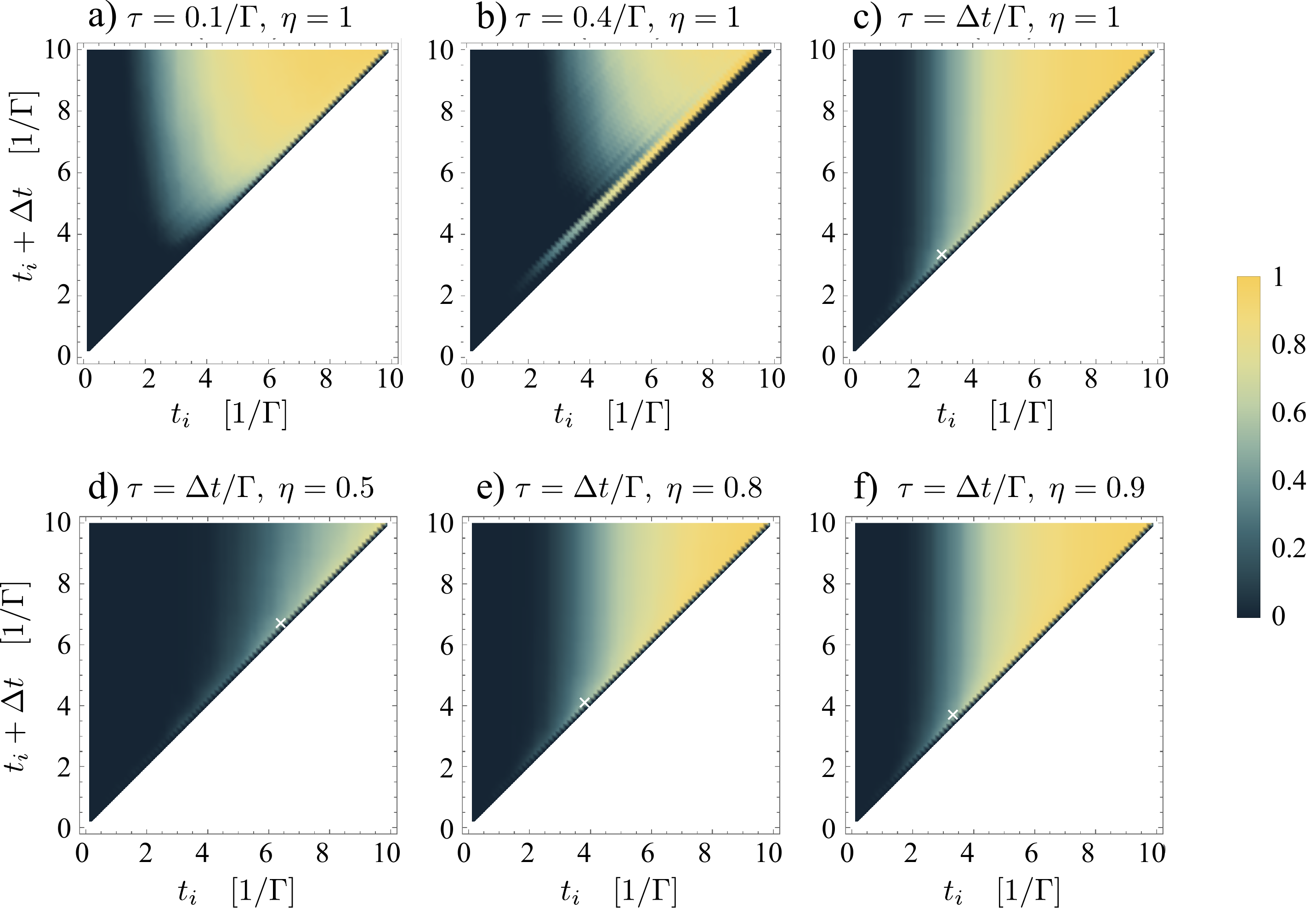} 
\end{center}
\caption{Success rate of the energetic-based estimator $\mathcal{E}_{ss}$ for different coarse-graining times, $\tau = 0.1/\Gamma$ (a), $\tau =0.4/\Gamma$ (b), $\tau = \Delta t/\Gamma$ (c), and for $\tau = \Delta t/\Gamma$ and different values of the detection efficiency $\eta = 1$ (a-b-c), $\eta = 0.5$ (d), $\eta = 0.8$ (e), $\eta = 0.9$ (f). The density plots show the success rate of $\mathcal{E}_{ss}$ for all integration intervals $[t_i, t_i + \Delta t]$, $t_i$ and $\Delta t$ ranging from $0$ to $10$. In the simulation, we considered as entangled the trajectories leading to a final concurrence greater or equal to $0.8$. The success rate is defined as the proportion of entangled trajectories for which the estimator takes a negative value over all entangled trajectories. The white crosses on plots (c-f) indicate the points of minimum $t_i$ reaching a success rate of $50\%$, i.e $\tau = \Delta t = 0.3/\Gamma$ and $t_i= 3/\Gamma$,  $6.4/\Gamma$,  $3.8/\Gamma$ and  $3.4/\Gamma$, respectively, for plots c), d), e) and f).}
\label{fig:esti}
\end{figure*} 

Inequalities \eqref{eq:ineq1} and \eqref{eq:ineq2} constitute one of the main analytical results of this work. The entanglement rate is exclusively upper and lower bounded by energetic quantities, the quantum heat and quantum heat fluctuations defined in Eq.\eqref{eq:tilde}. All quantities are defined over a (small) finite time interval $dt$. Remarkably, we can further exploit Ineq. \eqref{eq:ineq1} towards the derivation of an energetic-based estimator to assess the presence of entanglement at intermediate times. Of interest towards %non-equilibrium quantum thermodynamics
single shot entanglement detection, this will be done at the level of a unique quantum trajectory.

%  and depend implicitly on the measurement outcome $J_\gamma$. This inequality constitutes one of the main results of this work, together with the witnesses derived in the following section. 
%Of interest towards non-equilibrium quantum thermodynamics, these bounds are valid at the level of a unique stochastic (quantum) trajectory. This validity is illustrated in Fig.\ref{f:fig4} for two examples of trajectories, leading the qubits either into an entangled state or into a separable state. For the sake of clarity, we show the thermodynamical-based bounds with their fluctuations, whereas the concurrence rate has been averaged over a time interval of duration $0.4 / \Gamma$ (reflected into the error bars). In contrast to the lower bound for trajectories ending in a separable state that stays negative at long times, the lower bound for trajectories ending up in an entangled state rapidly decreases to 0 and remains positive and close to 0 at long times $t \gg 1/\Gamma$. This long-time bounds can be derived analytically: the lower bound for trajectories corresponding to qubits in an entangled state is trivially 0, whereas it reaches $-4 \Gamma$ for qubits ending up in a separable state, see Appendix \ref{appen:2}. These two limits are clearly distinguishable in Fig.\ref{f:fig4}.

\section{Single-shot energetic-based estimator for entanglement}
\label{sec5}

Formally, a witness for entanglement is an observable that takes a negative value when averaged with respect to a state that is entangled. If the witness takes a positive value, no conclusions can be drawn, the state can either be entangled or separable \cite{Plenio07}. In the past years, few witnesses based on temperature have been proposed, as a first attempt to exploit energetic quantities to certify the presence of entanglement \cite{Anders06, Anders08}. Following the spirit of assessing the presence of entanglement with some independent quantities, we introduce in this section a \emph{trajectory-based witness}, i.e. a quantity which takes negative value in presence of entanglement, just as usual entanglement witnesses, but which takes as an input a single quantum trajectory, or equivalently a weak measurement record, instead of a quantum state. As a first step to build this witness, we use the property that a given quantum trajectory $\gamma$ will drive the qubits onto an entangled state if,
\bb
\left\langle \!\!\!\left\langle \frac{d \mathcal{C}_\gamma(t')}{dt} \right\rangle \!\!\! \right\rangle_{dt}  \geq 0 \,,\quad\forall t' \in [0,t].\label{condition}
\ee
This condition is motivated by the idea that accumulating positive time-derivatives would lead to a positive concurrence at final time $t$. While one cannot certify the presence of entanglement along individual trajectory $\gamma$ from Eq.\eqref{condition}, the integration over time of $\frac{d \mathcal{C}_\gamma(t)}{dt}$ used to compute the concurrence at time $t$ is expected to play the role of the ensemble average $\left\langle \!\!\!\left\langle \cdot \right\rangle \!\!\! \right\rangle_{dt}$ for $\Gamma dt \ll 1$ and provided $t \gtrsim \Gamma^{-1}$. Therefore, condition \eqref{condition} implies an entangled state at time $t$ for a high fraction of the considered trajectories.

Making use of inequality \eqref{eq:ineq1}, this condition translates into:
\bb
\label{eq:lower_bound}
&&\tilde{\sigma}_\gamma(t) - 2 \tilde{Q}_\gamma^2(t) \mathcal{C}_\gamma(t) \geq 0\,,
\ee
We can now use that the concurrence $\mathcal{C}_\gamma$ takes values within $[0,1]$ to prove an inequality assessing the presence of entanglement, that only depends on energetic quantities, $\tilde{\sigma}_\gamma(t) - 2 \tilde{Q}_\gamma^2(t) \geq 0\,.$
%\bb
%\label{eq:lower_bound}
%\tilde{\sigma}_\gamma(t) - 2 \tilde{Q}_\gamma^2(t) \geq 0\,.
%&&
%\ee
We can now directly use this condition to introduce our trajectory-based entanglement witness $\mathcal{W}_\gamma$, defined as:
\bb
\label{eq:wit}
\mathcal{W}_\gamma =  \frac{1}{\Delta t}\int_{t_i}^{t_i+\Delta t} \left[2 \tilde{Q}_\gamma^2(t) - \tilde{\sigma}_\gamma(t) \right]dt \,.
\ee
The time-averaged over $\Delta t$ is meant to take into account a finite acquisition time during the experiment.
When {$\mathcal{W}_\gamma$} is negative, one can expect with high probability that the trajectory $\gamma$ leads to qubits in an entangled state. When it is positive, one can not draw a definite conclusion. However, and as stated before, this quantity implies an ensemble-average $\langle\!\langle \cdot \rangle\!\rangle_{dt}$ over all trajectories occurring during the finite time interval $dt$, {which is not yet optimal for experimental purposes}. We {therefore} define an estimator $\mathcal{E}_{ss}$ (with the label $ss$ referring to single-shot), where the ensemble average in the definition of $\sigma_\gamma(t)$ Eq.\eqref{eq:s_gamma} is replaced by a time average over an interval $\tau$. This procedure is similar to a coarse-graining of the fluctuations along a single trajectory. The witness \eqref{eq:wit} then transforms into an estimator, valid at the level of single trajectory:
%\bb
%2 \tilde{Q}_\gamma^2(t) - \tilde{\sigma}_\gamma(t) \Rightarrow 2 \tilde{Q}_\gamma(t)^2 \,dt -\sqrt{\frac{1}{\tau}\int_{t_i}^{t_i+\tau} \delta\tilde{Q}_\gamma^2(t) \,dt }
%\ee

\bb
\label{eq:E_ss}
\mathcal{E}_{ss} \!\!= \!\! \frac{1}{\Delta t}\int_{t_i}^{t_i+\Delta t} \left[2 \tilde{Q}_\gamma(t)^2  -\sqrt{\frac{1}{\tau}\int_{t}^{t+\tau} \delta\tilde{Q}_\gamma^2(t')\, dt' } \right]\!dt\,. \nonumber \\
\ee

%\deleted{The coarse-grained quantity (in square brackets) is then overall integrated over another finite time interval $\tau$ to satisfy potential experimental constraints. Experimentally, the usefulness of this estimator is as follows. Given a trajectory $\gamma$ defined by the set of measurement records $\{I_\gamma(t')\}_{t'<t}$, $\mathcal{E}_{ss}$ will take negative values at each instant of time $t$ if the energetic quantities \deleted{$\tilde{\sigma}_\gamma(t)$} \gh{$\delta\tilde{Q}_\gamma^2(t)$} and $\tilde{Q}_\gamma(t)$ during the time interval $dt$ satisfy \gh{the similar equation to }Eq.\eqref{eq:lower_bound}. Even at intermediate times (when the measurement is not yet projective), the fluctuations in the quantum heat provide a privileged access to the amount of entanglement between the qubits.} 
The performance of $\mathcal{E}_{ss}$ as an estimator to attest entanglement between the qubits is analyzed through two figures of merit, its success rate (ratio of detected vs. total number of trajectories leading to entangled states) and its error rate (whenever $\mathcal{E}_{ss}$ takes a negative value whereas the trajectory does not lead to entangled qubits). The error rate takes the maximal value of $0.2\%$ for the coarse-graining times $\tau = 0.1$ and $\tau = 0.4$ and $1.2\%$ for $\tau= \Delta t$. Indeed, as the coarse-graining time increases, the time average leads to trusty measurements, as expected from the analytical bound derived with the ensemble-averaged. The finite error rate forbids us to claim an energetic witness, but its small value demonstrates the usefulness of our energetic-based estimator. As shown in Fig.\ref{fig:esti}, the success rate of $\mathcal{E}_{ss}$ does not strongly depend on the overall integration window $\Delta t$, but rather on the initial time $t_i$ from which the averages are performed. At large $t_i$, the success rate reaches 1 as expected from the convergence of each trajectory towards one of the eigenstate of the measurement operator $\hat{\Phi}$. However, the success rate exceeds 0.5 for $t_i$ as small as $3/\Gamma$ for $\tau = \Delta t = 0.3/\Gamma$, with a corresponding error rate of $0.4\%$, being promising for future experiments towards single-shot energetic-based estimators to certify the presence of entanglement. 
Finally, as a first step towards realistic implementation, we have also investigated the robustness of our estimator against finite detection efficiency. We have numerically simulated trajectories when the half-parity detection channel has an efficiency $\eta<1$, which leads to mixed state trajectories $\rho_\gamma(t)$ \cite{Jacobs06}, see App. \ref{appen:3} . As a consequence, the analytic justification for the estimator is not valid anymore as the formula for the concurrence has to be modified for mixed state\cite{Wootters98} and there is no analytic expression for the state conditioned to a given readout $J_\gamma$. However, a numerical treatment is possible, and it is straightforward to extend the definition \eqref{eq:E_ss} to mixed state trajectories, using the average heat increment along mixed state trajectory $\gamma$ defined as $\delta Q_\gamma(t) = \text{Tr}\{d\rho_\gamma(t)\hat H_{\cal S}\}$. The latter is related to the populations of the states $\ket{\uu}$ and $\ket{\dd}$ and of the even and odd subspaces in the same way as in Eq.\eqref{eq:dQ1}, with an additional overall factor $\sqrt{\eta}$. The generation of entanglement happens to be quite robust to finite detection efficiency, as witnessed by the experimental implementations \cite{Riste13,Roch14}. The success rate, computed from a sample of $1000$ numerically generated trajectories, is plotted for finite efficiency in Fig.~\ref{fig:esti}d)-f) for different values of $\eta$, showing that our method is robust against finite detection efficiency. High success rates (up to $80\%$ for $\Delta t = 10/\Gamma$) are predicted even for $\eta = 0.5$ which is the order of magnitude of the experimental conditions. The error rate remains smaller or of the order of $1\%$.

\section{Discussion}

In this work, we investigate the energy fluctuations associated with entanglement genesis during the paradigmatic half-parity measurement procedure. Not only this measurement generates both product states and entangled states, but it also constitutes an energy observable. Based on a quantum-trajectory approach, and making use of the framework provided by stochastic thermodynamics, we demonstrate that the generation of entanglement is closely related to the presence of quantum heat fluctuations induced by the stochasticity of the weak continuous measurement. We derive analytical upper and lower bounds for the entanglement genesis rate. We then exploit the lower bound to derive an estimator that is solely based on energetic observables and their fluctuations. We show that this energetic-based estimator is indeed able to attest the presence of entanglement at a finite time for a given quantum trajectory, with a tunable probability that depends on the total time-averaged window $\Delta t$. Remarkably, this single-shot estimator is valid at the level of a single trajectory and so does not require any ensemble average. Remarkably, the energetic-based estimator we proposed is robust to finite detection efficiency, reaching a success rate close to 1 for suitable integration times as discussed earlier.

These results do not aim at evaluating the energetic cost of creating quantum correlations, see for instance Refs.\cite{Huber15,Bruschi15, Friis16}, but rather at defining an energetic signature associated to their generation. Whereas the quantum heat is zero on average, the quantum fluctuations $\delta Q$ during a finite time interval $dt$ contain relevant information to attest the presence of entanglement.

Exact in the context of the half-parity measurement process which also plays the role of an energetic filter, our work fixes theoretical tools for thermodynamic analysis of the generation of quantum correlations and opens the way to develop witnesses based exclusively on the measurements of energetic quantities for quantum information purposes. This theoretical research is additionally motivated by recent experimental achievements in the emergent field of quantum caloritronics, aiming at controlling and measuring energetic observables like the heat current in various quantum circuits \cite{Giazotto12, Jezouin13, Gasparinetti15, Iftikhar16, Fornieri17, Banerjee17,Zanten16}. \\

\textbf{Acknowledgements} 
G.H. gratefully acknowledges discussions with J. Anders,  A. Jordan and I. Siddiqi. C.E. acknowledges the US Department of Energy grant No. DE-SC0017890. This research was supported in part by the National Science Foundation under Grant No. NSF PHY-1748958. G.H. thanks the Swiss National Science Foundation for support through the MHV grant 164466 and starting grant PRIMA PR00P2$\_$179748. This research was supported in part by the National Science Foundation under Grant No. NSF PHY17-48958.

\appendix

% \section{Methods}
 \section{Long-time limit formula between the concurrence derivative and infinitesimal heat fluctuations}
 \label{appen:1}
 
The fluctuations of the heat increment associated to the lose of coherences between the two parity subspaces are defined in a similar was as $\sigma_\gamma$: 
\bb
\sigma_\gamma^{(eo)}(t) &=& \sqrt{\langle\!\langle \delta Q_\gamma^{(eo)2} (t) \rangle\!\rangle_{dt} }\,. \\
&=& 2 \epsilon \sqrt{\Gamma dt} \frac{e^{-2 \Gamma t} \cosh{4 \Gamma t J_\gamma}}{\left(1+ e^{-2 \Gamma t} \cosh{4 \Gamma t J_\gamma}\right)^2}\,. \label{eq:sigma_eo}
\ee
We can compare them to the derivative of the concurrence averaged over realizations occurring during the time interval $dt$:
\bb
\label{eq:dC_eo1}
\left\langle\!\!\!\left\langle \frac{d \mathcal{C}_\gamma}{dt} \right\rangle\!\!\!\right\rangle_{dt} &=& 2 \Gamma \frac{e^{-2 \Gamma t} + e^{- 2 \Gamma t} \cosh(4 \Gamma t J_\gamma)}{(1+e^{-2 \Gamma t} \cosh(4 \Gamma t J_\gamma))^2} \nonumber \\
&&- 4 \Gamma \, \frac{Q_\gamma}{\epsilon} \, \mathcal{C}_\gamma  \moy{\hat\Phi(t)}_\gamma .
\ee
When $t \gg 1/\Gamma$, the probability distribution of the measurement outcome $J_\gamma$ is narrowly peaked around the three values corresponding to the eigenvalues of the half-parity measurement operator $\hat{\Phi}$ \cite{Jacobs06}, i.e. $\pm1,0$. The average value of $\hat{\Phi}$ tends also to one of the three eigenvalues. Hence, for trajectories corresponding to qubits in a maximally entangled state at long times, $J_\gamma = \moy{\hat\Phi(t)}_\gamma = 0$ and the heat flow $Q_\gamma = 0$. Consequently, Eqs. \eqref{eq:sigma_eo} and \eqref{eq:dC_eo1} simplify to
\bb
\left.{\sigma}_\gamma^{(eo)}\right\vert_{J_\gamma = 0} &=& 2 \epsilon \sqrt{\Gamma dt }  \frac{e^{-2 \Gamma t}}{(1+ e^{-2 \Gamma t})^2} \\
\left.\left\langle\!\!\!\left\langle \frac{d \mathcal{C}_\gamma}{dt}\right\rangle\!\!\!\right\rangle_{dt}\right\vert_{J_\gamma = 0} &=&  4 \Gamma  \frac{e^{-2 \Gamma t}}{(1+ e^{-2 \Gamma t})^2} \,.
\ee
 When time exceeds the measurement time, the following equality holds:
 \bb
 \left.\left\langle\!\!\!\left\langle \frac{d \mathcal{C}_\gamma}{dt} \right\rangle\!\!\!\right\rangle_{dt}\right\vert_{J_\gamma = 0} = \frac{2}{\epsilon} \sqrt{\frac{\Gamma}{dt}} \left.{\sigma}_\gamma^{(eo)} \right\vert_{J_\gamma = 0}\,.
\ee
Although only valid and meaningful at times longer than the measurement time, this relation exemplifies the underlying fundamental role of infinitesimal heat fluctuations for the generation of entanglement. As stated in the main text, this relation is only exact in the context of the half-parity measurement considered in this work and can not be exploited experimentally. Indeed, one could not distinguish in an experiment infinitesimal heat fluctuations originating in the loss of phase coherence between the two parity subspaces ($\sigma_\gamma^{(eo)}$) from total infinitesimal heat fluctuations ($\sigma_\gamma$).

 \section{Energetic bounds for the entanglement genesis rate}
 \label{appen:2}
 
 The general expression of the concurrence derivative reads
 \bb
 \frac{d \mathcal{C}_\gamma}{dt} &=& 2 \Gamma \frac{e^{-2 \Gamma t} + e^{- 2 \Gamma t} \cosh(4 \Gamma t J_\gamma)}{(1+e^{-2 \Gamma t} \cosh(4 \Gamma t J_\gamma))^2} \nonumber \\
 &&- 4 \Gamma \frac{e^{-2 \Gamma t} \sinh(4 \Gamma t J_\gamma) (1- e^{-2 \Gamma t}) d(J_\gamma t)/dt}{(1+e^{-2 \Gamma t} \cosh(4 \Gamma t J_\gamma))^2} \nonumber \\
 &&
 \ee
Using Eqs. \eqref{Jgamma},\eqref{eq:C} and \eqref{eq:Q}, it can be rewritten as
\bb
 \frac{d \mathcal{C}_\gamma}{dt} &=& 2 \Gamma \frac{e^{-2 \Gamma t} + e^{- 2 \Gamma t} \cosh(4 \Gamma t J_\gamma)}{(1+e^{-2 \Gamma t} \cosh(4 \Gamma t J_\gamma))^2}\nonumber \\
 && - 4 \Gamma \, \frac{Q_\gamma}{\epsilon} \, \mathcal{C}_\gamma  \left(\moy{\hat\Phi(t)}_\gamma + \dfrac{dW_\gamma(t)}{dt}\right).
\ee
To enable the comparison with $\tilde{\sigma}_\gamma$, we perform the ensemble average $\langle\!\langle \cdot \rangle\!\rangle_{dt}$ and obtain:
\bb
\left\langle\!\!\!\left\langle \frac{d \mathcal{C}_\gamma}{dt} \right\rangle\!\!\!\right\rangle_{dt} &=& 2 \Gamma \frac{e^{-2 \Gamma t} + e^{- 2 \Gamma t} \cosh(4 \Gamma t J_\gamma)}{(1+e^{-2 \Gamma t} \cosh(4 \Gamma t J_\gamma))^2} \nonumber \\
&&- 4 \Gamma \, \frac{Q_\gamma}{\epsilon} \, \mathcal{C}_\gamma  \moy{\hat\Phi(t)}_\gamma\,,
\ee
which corresponds to Eq.\eqref{eq:dC_eo} in the main text.

Using the inequalities
\bb
&&e^{-2 \Gamma t} + e^{- 2 \Gamma t} \cosh(4 \Gamma t J_\gamma) \leq 2 e^{-2 \Gamma t} \cosh(4 \Gamma t J_\gamma)  \,, \nonumber \\
&&
\ee
and 
\bb
&&e^{-2 \Gamma t} + e^{- 2 \Gamma t} \cosh(4 \Gamma t J_\gamma)  \geq e^{-4 \Gamma t} + e^{- 2 \Gamma t} \cosh(4 \Gamma t J_\gamma) \,, \nonumber \\
&&
\ee
we can now compare the r.h.s. of these two inequalities with the st. dev. of the total heat fluctuations $\tilde{\sigma}_\gamma(t)$:
\bb
\tilde{\sigma}_\gamma(t) = \frac{e^{-4 \Gamma t} + e^{- 2 \Gamma t} \cosh(4 \Gamma t J_\gamma) }{(1+e^{-2 \Gamma t} \cosh(4 \Gamma t J_\gamma))^2}\,.
\ee
These bounds directly lead to Eqs. \eqref{eq:ineq1} and \eqref{eq:ineq2} in the main text. \\

 \section{Finite detection efficiency}
 \label{appen:3}

When the efficiency of the detector takes a finite value $\eta$, the state of the two qubits along a given trajectory $\gamma$ is a mixed state $\rho_\gamma(t)$ which obeys \cite{Jacobs06}:
\bb
&&d\rho_\gamma(t) = -idt[\hat H_{\cal S},\rho_\gamma(t)]\nonumber\\
&& + \frac{\Gamma dt}{2}\left(\hat\Phi\rho_\gamma(t)\hat\Phi - \tfrac{1}{2}\{\hat\Phi^2,\rho_\gamma(t)\}\right)\nonumber\\
&& + \sqrt{\eta \Gamma}dW_\gamma(t)\left(\{\hat\Phi,\rho_\gamma(t)\} - 2\moy{\hat\Phi(t)}\rho_\gamma(t\right),
\ee
where $\{A,B\} = AB+BA$. The measurement record $I_\gamma(t)$ is now linked to the Wiener increment $dW_\gamma(t)$ via:
\bb
I_\gamma(t) = \moy{\hat\Phi(t)}_\gamma + \frac{dW_\gamma(t)}{2\sqrt{\eta\Gamma}dt}.
\ee

\bibliographystyle{apsrev4-1_modified}
\bibliography{parity_thermo}

\end{document}